\documentstyle[prl,aps]{revtex}

\begin{document}
\author{J.Q. Shen}
\address{Zhejiang Institute of Modern Physics and Department of Physics,
Zhejiang University, Hangzhou 310027, P.R. China}
\date{\today }
\title{The interaction between graviton spin and gravitomagnetic fields\footnote{This paper presenting the trivial and lengthy calculations only serves as the supplement to a paper entitled ``The purely gravitational generalization of spin-rotation couplings'' (by J.Q. Shen).}}
\maketitle

\begin{abstract}
This note is devoted to the detailed mathematical treatment of the
coupling of graviton spin to gravitomagnetic fields. The
expression ({\it i.e.}, $\sim
 g_{0m}\dot{g}_{0n}(\partial_{m}g_{0n}-\partial_{n}g_{0m})$) for
the graviton spin-gravitomagnetic (S-G) coupling in the
Lagrangian/Hamiltonian density of the weak gravitational fields is
presented in this note.


\end{abstract}

The interaction of graviton spin with gravitomagnetic fields under
consideration is the extension of Mashhoon's spin-rotation
coupling\cite{Mashhoon2}, which is the interaction between the
gravitomagnetic moment of a spinning particle and the noninertial
frame of reference. To the best of our knowledge, the
spin-rotation coupling of photon, electron and neutron has been
taken into account in the literature\cite{Mashhoon2,Shen}.
However, the gravitational coupling of graviton spin to the
gravitomagnetic fields, which may be of physical interest, has so
far never yet been considered. In this note, we will extend
Mashhoon's spin-rotation coupling to a purely gravitational case,
where the graviton spin will be coupled to gravitomagnetic fields.
Here we are concerned mainly with the detailed mathematical
treatment of the coupling of graviton spin to gravitomagnetic
fields.

First let us consider the Christoffel symbol of the weak
gravitational field, the first- and second- order terms of which
may be expressed as follows\cite{Shao1}
\begin{eqnarray}
\Gamma^{\alpha}_{\ \beta
\gamma}&=&-\frac{1}{2}K\left(h^{\alpha}_{\beta,\gamma}+h^{\alpha}_{\gamma,\beta}-h^{,\alpha}_{\beta\gamma}-\frac{1}{2}\delta^{\alpha}_{\beta}h^{\lambda}_{\lambda,\gamma}-\frac{1}{2}\delta^{\alpha}_{\gamma}h^{\lambda}_{\lambda,\beta}+\frac{1}{2}\eta_{\beta\gamma}h^{\lambda,\alpha}_{\lambda}\right)
\nonumber \\
&+&\frac{1}{2}K^{2}\left(h_{\beta\lambda}h^{\lambda\alpha}_{,\gamma}+h_{\gamma\lambda}h^{\lambda\alpha}_{,\beta}+h^{\alpha\lambda}h_{\beta\gamma,\lambda}-h_{\beta\lambda}h^{\lambda,\alpha}_{\gamma}-h_{\gamma\lambda}h^{\lambda,\alpha}_{\beta}\right)
\nonumber \\
&+&\frac{1}{2}K^{2}\left(-\frac{1}{2}\delta^{\alpha}_{\beta}h_{\lambda\tau}h^{\lambda\tau}_{,\gamma}-\frac{1}{2}\delta^{\alpha}_{\gamma}h_{\lambda\tau}h^{\lambda\tau}_{,\beta}-\frac{1}{2}\eta_{\beta\gamma}h^{\alpha\lambda}h^{\tau}_{\tau,\lambda}+\frac{1}{2}h_{\beta\gamma}h^{\lambda,\alpha}_{\lambda}+\frac{1}{2}\eta_{\beta\gamma}h_{\lambda\tau}h^{\lambda\tau,\alpha}\right)+{\mathcal
O}(h^{3}),
\end{eqnarray}
where $K$ and $h^{\mu\nu}$ are so defined that
$\sqrt{-g}g^{\mu\nu}=\eta^{\mu\nu}+Kh^{\mu\nu}$ is satisfied.
Since there exists an exact analogy between general relativity and
electrodynamics for weak gravitational
fields\cite{Mashhoon,Harris}, one may think of the expression
associated with
$g_{0m}\dot{g}_{0n}(\partial_{m}g_{0n}-\partial_{n}g_{0m})$ in the
gravitational Hamiltonian/Lagrangian density as the interaction
term of the graviton spin with the gravitomagnetic fields, where
dot denotes the time derivative of ${g}_{0n}$. If the Lagrangian
density ${\mathcal L}_{\rm s-g}$ contains the expression $\sim
g_{0m}\dot{g}_{0n}(\partial_{m}g_{0n}-\partial_{n}g_{0m})$ (here
the both indices $m$ and $n$ take $1, 2, 3$, {\it i.e.}, summation
is carried out over $1, 2, 3$), then the Hamiltonian density of
the weak gravitational field may be of the form
$T^{00}=2\frac{\partial {\mathcal L}_{\rm s-g}}{\partial
\dot{g}_{0n}}\dot{g}_{0n}-{\mathcal L}_{\rm s-g}={\mathcal L}_{\rm
s-g}$. It should be noted that here the factor $2$ results from
the fact that $\frac{\partial {\mathcal L}_{\rm s-g}}{\partial
\dot{g}_{\mu\nu}}\dot{g}_{\mu\nu}$ includes $\frac{\partial
{\mathcal L}_{\rm s-g}}{\partial
\dot{g}_{0n}}\dot{g}_{0n}+\frac{\partial {\mathcal L}_{\rm
s-g}}{\partial \dot{g}_{n0}}\dot{g}_{n0}$. So, in order to find
the interaction term between the graviton spin and the
gravitomagnetic fields (in what follows it will be referred to as
the graviton S-G coupling term) in the Hamiltonian density of the
weak gravitational field, we should only take into consideration
the graviton S-G coupling term in the Lagrangian density ({\it
i.e.}, $\sqrt{-g}g^{\mu\nu}R_{\mu\nu}$). Because of
$g^{\mu\nu}R_{\mu\nu}$ contains $g^{0m}R_{0m}+g^{m0}R_{m0}$ ({\it
i.e.}, $2g^{0m}R_{0m}$), we will analyze the $0m$ component of the
Ricci tensor $R_{\mu\nu}$ in the following. Note that the
expression for $R_{\mu\nu}$ is written as
$R_{\mu\nu}=\Gamma^{\rho}_{\ \mu\rho,\nu}-\Gamma^{\rho}_{\
\mu\nu,\rho}-\Gamma^{\sigma}_{\ \mu\nu}\Gamma^{\rho}_{\
\sigma\rho}+\Gamma^{\rho}_{\ \sigma\nu}\Gamma^{\sigma}_{\
\mu\rho}$. Now we will extract the expression $\sim
\dot{g}_{0n}(\partial_{m}g_{0n}-\partial_{n}g_{0m})$ from
$R_{\mu\nu}$.

First we consider the terms $\Gamma^{\rho}_{\
\mu\rho,\nu}-\Gamma^{\rho}_{\ \mu\nu,\rho}$ in $R_{\mu\nu}$.

\subsection{$\Gamma^{\rho}_{\ \mu\rho,\nu}-\Gamma^{\rho}_{\ \mu\nu,\rho}$}
The second-order term (the coefficient of which is
$\frac{1}{2}K^{2}$) in $\Gamma^{\rho}_{\ \mu\rho}$ reads
\begin{eqnarray}
\Gamma^{\rho}_{\mu\rho}\left(\propto
\frac{1}{2}K^{2}\right)&=&h_{\mu\lambda}h^{\lambda\rho}_{,\rho}+h_{\rho\lambda}h^{\lambda\rho}_{,\mu}+h^{\rho\lambda}h_{\mu\rho,\lambda}-h_{\mu\lambda}h^{\lambda,\rho}_{\rho}-h_{\rho\lambda}h^{\lambda,\rho}_{\mu}
\nonumber \\
&-&\frac{1}{2}h_{\lambda\tau}h^{\lambda\tau}_{,\mu}-2h_{\lambda\tau}h^{\lambda\tau}_{,\mu}-\frac{1}{2}h^{\lambda}_{\mu}h^{\tau}_{\tau,\lambda}+\frac{1}{2}h_{\mu\rho}h^{\lambda,\rho}_{\lambda}+\frac{1}{2}h_{\lambda\tau}h^{\lambda\tau}_{,\mu}.
\end{eqnarray}
Thus the terms, which will probably contribute to the graviton S-G
coupling term, in the derivative of $\Gamma^{\rho}_{\ \mu\rho}$
with respect to $x^{\nu}$, {\it i.e.}, $\Gamma^{\rho}_{\
\mu\rho,\nu}$, are given as follows
\begin{equation}
h_{\mu\lambda,\nu}h^{\lambda\rho}_{,\rho}+h_{\rho\lambda,\nu}h^{\lambda\rho}_{,\mu}+h^{\rho\lambda}_{,\nu}h_{\mu\rho,\lambda}-h_{\mu\lambda,\nu}h^{\lambda,\rho}_{\rho}-h_{\rho\lambda,\nu}h^{\lambda,\rho}_{\mu}-\frac{5}{2}h_{\lambda\tau,\nu}h^{\lambda\tau}_{,\mu}-\frac{1}{2}h^{\lambda}_{\mu,\nu}h^{\tau}_{\tau,\lambda}+\frac{1}{2}h_{\mu\rho,\nu}h^{\lambda,\rho}_{\lambda}+\frac{1}
{2}h_{\lambda\tau,\nu}h^{\lambda\tau}_{,\mu}.
\end{equation}
But via the detailed analysis we find that the terms that truly
gives contribution to the graviton S-G coupling are
$-h_{\rho\lambda,\nu}h^{\lambda\rho}_{,\mu}$ and
$h_{\rho\lambda,\nu}\left(h_{\mu}^{\rho,\lambda}-h_{\mu}^{\lambda,\rho}\right)$
only, the latter of which, however, vanishes. It is clearly seen
that
$-\sqrt{-g}g^{\mu\nu}h_{\rho\lambda,\nu}h^{\lambda\rho}_{,\mu}$
contains
$-\sqrt{-g}g^{0m}h_{\rho\lambda,0}h^{\lambda\rho}_{,m}-\sqrt{-g}g^{0m}h_{\rho\lambda,m}h^{\lambda\rho}_{,0}$
({\it i.e.},
$-2\sqrt{-g}g^{0m}h_{\rho\lambda,0}h^{\lambda\rho}_{,m}$), which
contains
\begin{equation}
-2\sqrt{-g}g^{0m}h_{0n,0}h^{n0}_{,m}-2\sqrt{-g}g^{0m}h_{n0,0}h^{0n}_{,m}=-4\sqrt{-g}g^{0m}\dot{g}_{n}h^{n0}_{,m}=4\sqrt{-g}g^{0m}\dot{g}_{n}\partial_{m}g_{n}.
\label{eq40}
\end{equation}
Here $h^{0n}=g^{n}=-g_{n}$. The flat Minkowski metric
$\eta_{\mu\nu}={\rm diag}[+1, -1,-1,-1]$. Clearly, the relation
between the metric $g^{0m}$ (with $m=1,2,3$) and the
gravitomagnetic vector potential $g^{m}$ is that
$\sqrt{-g}g^{0m}=Kh^{0m}=Kg^{m}$.
\\ \\

The second-order term (the coefficient of which is
$\frac{1}{2}K^{2}$) in $\Gamma^{\rho}_{\ \mu\nu}$ reads

\begin{eqnarray}
\Gamma^{\rho}_{\ \mu\nu}\left(\propto
\frac{1}{2}K^{2}\right)&=&h_{\mu\lambda}h^{\lambda\rho}_{,\nu}+h_{\nu\lambda}h^{\lambda\rho}_{,\mu}+h^{\rho\lambda}h_{\mu\nu,\lambda}-h_{\mu\lambda}h^{\lambda,\rho}_{\nu}-h_{\nu\lambda}h^{\lambda,\rho}_{\mu}                 \nonumber \\
&-&\frac{1}{2}\delta^{\rho}_{\mu}h_{\lambda\tau}h^{\lambda\tau}_{,\nu}-\frac{1}{2}\delta^{\rho}_{\nu}h_{\lambda\tau}h^{\lambda\tau}_{,\mu}-\frac{1}{2}\eta_{\mu\nu}h^{\rho\lambda}h^{\tau}_{\tau,\lambda}+\frac{1}{2}h_{\mu\nu}h^{\lambda,\rho}_{\lambda}+\frac{1}{2}\eta_{\mu\nu}h_{\lambda\tau}h^{\lambda\tau,\rho}.
\end{eqnarray}
Those terms in $\Gamma^{\rho}_{\ \mu\nu,\rho}$ which may have
effect on the graviton S-G coupling are
\begin{eqnarray}
\left(h_{\mu\lambda,\rho}h^{\lambda\rho}_{,\nu}+h_{\nu\lambda,\rho}h^{\lambda\rho}_{,\mu}\right)+h^{\rho\lambda}_{,\rho}h_{\mu\nu,\lambda}-\left(h_{\mu\lambda,\rho}h^{\lambda,\rho}_{\nu}+h_{\nu\lambda,\rho}h^{\lambda,\rho}_{\mu}\right)-\frac{1}{2}\left(h_{\lambda\tau,\mu}h^{\lambda\tau}_{,\nu}+h_{\lambda\tau,\nu}h^{\lambda\tau}_{,\mu}\right)
\nonumber  \\
-\frac{1}{2}\eta_{\mu\nu}h^{\rho\lambda}_{,\nu}h^{\tau}_{\tau,\lambda}+\frac{1}{2}h_{\mu\nu,\rho}h^{\lambda,\rho}_{\lambda}+\frac{1}{2}\eta_{\mu\nu}h_{\lambda\tau,\rho}h^{\lambda\tau,\rho}.
\end{eqnarray}
Further analysis shows that the terms which truly give
contribution to the graviton S-G coupling are the following four
terms:

(i)
$\sqrt{-g}g^{\mu\nu}\left(h_{\mu\lambda,\rho}h^{\lambda\rho}_{,\nu}+h_{\nu\lambda,\rho}h^{\lambda\rho}_{,\mu}\right)$
contains
\begin{equation}
\sqrt{-g}g^{0m}\left(h_{0\lambda,\rho}h^{\lambda\rho}_{,m}+h_{m\lambda,\rho}h^{\lambda\rho}_{,0}+h_{0\lambda,\rho}h^{\lambda\rho}_{,m}+h_{m\lambda,\rho}h^{\lambda\rho}_{,0}\right)=2\sqrt{-g}g^{0m}\left(h_{0\lambda,\rho}h^{\lambda\rho}_{,m}+h_{m\lambda,\rho}h^{\lambda\rho}_{,0}\right),
\end{equation}
which includes the following terms
\begin{equation}
2\sqrt{-g}g^{0m}\left(h_{0n,0}h^{n0}_{,m}+h_{m0,n}h^{0n}_{,0}\right)=-2\sqrt{-g}g^{0m}\dot{g}_{n}\left(\partial_{m}g_{n}+\partial_{n}g_{m}\right).
\label{eq1}
\end{equation}

(ii)
$-\sqrt{-g}g^{\mu\nu}\left(h_{\mu\lambda,\rho}h^{\lambda,\rho}_{\nu}+h_{\nu\lambda,\rho}h^{\lambda,\rho}_{\mu}\right)$
contains
\begin{equation}
-\sqrt{-g}g^{0m}\left(h_{0\lambda,\rho}h^{\lambda,\rho}_{m}+h_{m\lambda,\rho}h^{\lambda,\rho}_{0}+h_{0\lambda,\rho}h^{\lambda,\rho}_{m}+h_{m\lambda,\rho}h^{\lambda,\rho}_{0}\right)=-2\sqrt{-g}g^{0m}\left(h_{0\lambda,\rho}h^{\lambda,\rho}_{m}+h_{m\lambda,\rho}h^{\lambda,\rho}_{0}\right),
\end{equation}
which will give no contribution to the graviton S-G coupling. So
here we will not further consider it.

(iii)
$\sqrt{-g}g^{\mu\nu}h^{\rho\lambda}_{,\rho}h_{\mu\nu,\lambda}$
contains
\begin{equation}
\sqrt{-g}g^{0m}\left(h^{\rho\lambda}_{,\rho}h_{0m,\lambda}+h^{\rho\lambda}_{,\rho}h_{m0,\lambda}\right)=2\sqrt{-g}g^{0m}h^{\rho\lambda}_{,\rho}h_{0m,\lambda},
\end{equation}
which includes
\begin{equation}
2\sqrt{-g}g^{0m}h^{0n}_{,0}h_{0m,n}=-2\sqrt{-g}g^{0m}\dot{g}_{n}\partial_{n}g_{m}.
\label{eq3}
\end{equation}

(iv)
$-\frac{1}{2}\sqrt{-g}g^{\mu\nu}\left(h_{\lambda\tau,\mu}h^{\lambda\tau}_{,\nu}+h_{\lambda\tau,\nu}h^{\lambda\tau}_{,\mu}\right)$
({\it i.e.},
$-\sqrt{-g}g^{\mu\nu}h_{\lambda\tau,\mu}h^{\lambda\tau}_{,\nu}$)
includes
\begin{equation}
\sqrt{-g}g^{0m}\left(-h_{\lambda\tau,0}h^{\lambda\tau}_{,m}-h_{\lambda\tau,m}h^{\lambda\tau}_{,0}\right)=-2\sqrt{-g}g^{0m}h_{\lambda\tau,0}h^{\lambda\tau}_{,m},
\end{equation}
which contains
\begin{equation}
\sqrt{-g}g^{0m}\left(-2h_{0n,0}h^{0n}_{,m}-2h_{n0,0}h^{n0}_{,m}\right)=-4\sqrt{-g}g^{0m}h_{0n,0}h^{0n}_{,m}=4\sqrt{-g}g^{0m}\dot{g}_{n}\partial_{m}g_{n}.
\label{eq4}
\end{equation}

Thus it follows from Eq.(\ref{eq1}), (\ref{eq3}) and (\ref{eq4})
that the terms in $\sqrt{-g}g^{\mu\nu}\Gamma^{\rho}_{\
\mu\nu,\rho}$, which contribute to the graviton S-G coupling, are
given as follows
\begin{equation}
\sqrt{-g}g^{0m}\left[-2\dot{g}_{n}\left(\partial_{m}g_{n}+\partial_{n}g_{m}\right)-2\dot{g}_{n}\partial_{n}g_{m}+4\dot{g}_{n}\partial_{m}g_{n}\right]=\sqrt{-g}g^{0m}\left[2\dot{g}_{n}\partial_{m}g_{n}-4\dot{g}_{n}\partial_{n}g_{m}\right].
\label{eq140}
\end{equation}

{\bf THE RESULT}:

Hence it follows from (\ref{eq40}) and (\ref{eq140}) that the
terms in $\sqrt{-g}g^{\mu\nu}\left(\Gamma^{\rho}_{\
\mu\rho,\nu}-\Gamma^{\rho}_{\ \mu\nu,\rho}\right)$ which
contribute to the graviton S-G coupling are written in the form
\begin{equation}
\sqrt{-g}g^{0m}\left[4\dot{g}_{n}\partial_{m}g_{n}-\left(2\dot{g}_{n}\partial_{m}g_{n}-4\dot{g}_{n}\partial_{n}g_{m}\right)\right]=\sqrt{-g}g^{0m}\left(2\dot{g}_{n}\partial_{m}g_{n}+4\dot{g}_{n}\partial_{n}g_{m}\right).
\label{15}
\end{equation}

\subsection{$-\Gamma^{\sigma}_{\ \mu\nu}\Gamma^{\rho}_{\
\sigma\rho}$}

The first-order terms (proportional to $-\frac{1}{2}K$) in
$\Gamma^{\sigma}_{\ \mu\nu}$ is of the form
\begin{equation}
\Gamma^{\sigma}_{\ \mu\nu}\left(\propto
-\frac{1}{2}K\right)=h^{\sigma}_{\mu,\nu}+h^{\sigma}_{\nu,\mu}-h^{,\sigma}_{\mu\nu}-\frac{1}{2}\delta^{\sigma}_{\mu}h^{\lambda}_{\lambda,\nu}-\frac{1}{2}\delta^{\sigma}_{\nu}h^{\lambda}_{\lambda,\mu}+\frac{1}{2}\eta_{\mu\nu}h^{\lambda,\sigma}_{\lambda},
\end{equation}
which includes the {\it valuable} terms
$-\frac{1}{2}K\left(h^{\sigma}_{\mu,\nu}+h^{\sigma}_{\nu,\mu}-h^{,\sigma}_{\mu\nu}\right)$.
In the meanwhile, the terms proportional to $-\frac{1}{2}K$ in
$\Gamma^{\rho}_{\ \sigma\rho}$ take the following form
\begin{equation}
\Gamma^{\rho}_{\ \sigma\rho}\left(\propto
-\frac{1}{2}K\right)=-\frac{1}{2}K\left(h^{\rho}_{\sigma,\rho}+h^{\rho}_{\rho,\sigma}-h^{,\rho}_{\sigma\rho}+...\right)=-\frac{1}{2}Kh^{\rho}_{\rho,\sigma}.
\end{equation}

It is readily verified that the contribution of
$-\Gamma^{\sigma}_{\ \mu\nu}\Gamma^{\rho}_{\ \sigma\rho}$ to the
graviton S-G coupling is vanishing. So, we will not consider it
further in this note.

\subsection{$\Gamma^{\rho}_{\ \sigma\nu}\Gamma^{\sigma}_{\
\mu\rho}$}

Apparently, $\sqrt{-g}g^{\mu\nu}\Gamma^{\rho}_{\
\sigma\nu}\Gamma^{\sigma}_{\ \mu\rho}$ includes the following
terms
\begin{equation}
\sqrt{-g}g^{0m}\left(\Gamma^{\rho}_{\ \sigma0}\Gamma^{\sigma}_{\
m\rho}+\Gamma^{\rho}_{\ \sigma m}\Gamma^{\sigma}_{\
0\rho}\right)=\sqrt{-g}g^{0m}\left(\Gamma^{\rho}_{\
\sigma0}\Gamma^{\sigma}_{\ m\rho}+\Gamma^{\rho}_{\
0\sigma}\Gamma^{\sigma}_{\ \rho
m}\right)=2\sqrt{-g}g^{0m}\Gamma^{\rho}_{\
\sigma0}\Gamma^{\sigma}_{\ m\rho},
\end{equation}
where
\begin{equation}
\Gamma^{\rho}_{\ \sigma0}\left(\propto
-\frac{1}{2}K\right)=h^{\rho}_{\sigma,0}+h^{\rho}_{0,\sigma}-h^{,\rho}_{\sigma
0},
\end{equation}
and
\begin{equation}
\Gamma^{\sigma}_{\ m\rho}\left(\propto
-\frac{1}{2}K\right)=h^{\sigma}_{m,\rho}+h^{\sigma}_{\rho,
m}-h^{,\sigma}_{m\rho}.
\end{equation}
Thus, one can arrive at
\begin{eqnarray}
\left(h^{\rho}_{\sigma,0}+h^{\rho}_{0,\sigma}-h^{,\rho}_{\sigma
0}\right)\left(h^{\sigma}_{m,\rho}+h^{\sigma}_{\rho,
m}-h^{,\sigma}_{m\rho}\right)&=&\left[h^{\rho}_{\sigma,0}\left(h^{\sigma}_{\rho,
m}-h^{,\sigma}_{m\rho}\right)\right]+h^{\rho}_{\sigma,0}h^{\sigma}_{m,\rho}                \nonumber \\
&+&\left(h^{\rho}_{0,\sigma}-h^{,\rho}_{\sigma
0}\right)\left(h^{\sigma}_{m,\rho}+h^{\sigma}_{\rho,
m}-h^{,\sigma}_{m\rho}\right).             \label{eq177}
\end{eqnarray}
In the following discussions, for convenience, we classify the
terms on the right-handed side of Eq.(\ref{eq177}) into three
categories:

(i) $h^{\rho}_{\sigma,0}\left(h^{\sigma}_{\rho,
m}-h^{,\sigma}_{m\rho}\right)$ contains
\begin{equation}
h^{0}_{n,0}\left(h^{n}_{0,
m}-h^{,n}_{m0}\right)=-\dot{g}_{n}\left(\partial_{m}g_{n}-\partial_{n}g_{m}\right).
\label{eq22}
\end{equation}

(ii) $h^{\rho}_{\sigma,0}h^{\sigma}_{m,\rho}$ contains
\begin{equation}
h^{n}_{0,0}h^{0}_{m,n}=-\dot{g}_{n}\partial_{n}g_{m}.
\label{eq23}
\end{equation}

(iii) $\left(h^{\rho}_{0,\sigma}-h^{,\rho}_{\sigma
0}\right)\left(h^{\sigma}_{m,\rho}+h^{\sigma}_{\rho,
m}-h^{,\sigma}_{m\rho}\right)$ contains the following six terms
\begin{eqnarray}
h^{\rho}_{0,\sigma}h^{\sigma}_{m,\rho}&=&
h^{n}_{0,0}h^{0}_{m,n}+h^{0}_{0,0}h^{0}_{m,0}\sim -\dot{g}_{n}\partial_{n}g_{m},                 \nonumber \\
h^{\rho}_{0,\sigma}h^{\sigma}_{\rho, m}&=&h^{n}_{0,0}h^{0}_{n, m}+h^{0}_{0,0}h^{0}_{0, m}\sim -\dot{g}_{n}\partial_{m}g_{n},          \nonumber \\
-h^{\rho}_{0,\sigma}h^{,\sigma}_{m\rho} &=& ({\rm
giving \ no \ contribution \ to \ S-G \ coupling}),          \nonumber \\
-h^{,\rho}_{\sigma 0}h^{\sigma}_{m,\rho} &=&  ({\rm
giving \ no \ contribution \ to \ S-G \ coupling}),          \nonumber \\
-h^{,\rho}_{\sigma 0}h^{\sigma}_{\rho,m}&=&-h^{,0}_{n 0}h^{n}_{0,m}-h^{,0}_{0 0}h^{0}_{0,m}\sim \dot{g}_{n}\partial_{m}g_{n},     \nonumber \\
\left(-h^{,\rho}_{\sigma
0}\right)\left(-h^{,\sigma}_{m\rho}\right)&=&h_{n0}^{,0}h_{m0}^{,n}+h_{00}^{,0}h_{m0}^{,0}\sim
-\dot{g}_{n}\partial_{n}g_{m}.           \label{eq24}
\end{eqnarray}
So, the terms in $\left(h^{\rho}_{0,\sigma}-h^{,\rho}_{\sigma
0}\right)\left(h^{\sigma}_{m,\rho}+h^{\sigma}_{\rho,
m}-h^{,\sigma}_{m\rho}\right)$, which will have effect on the
graviton S-G coupling, are
\begin{equation}
-2\dot{g}_{n}\partial_{n}g_{m}.   \label{eq25}
\end{equation}

{\bf THE RESULT}:

Thus it follows from (\ref{eq22}), (\ref{eq23}) and (\ref{eq25})
that the total terms in $\sqrt{-g}g^{\mu\nu}\Gamma^{\rho}_{\
\sigma\nu}\Gamma^{\sigma}_{\ \mu\rho}$ which will give
contribution to the graviton S-G coupling are expressed as follows
\begin{equation}
2\sqrt{-g}g^{0m}\left[-\dot{g}_{n}\left(\partial_{m}g_{n}-\partial_{n}g_{m}\right)-\dot{g}_{n}\partial_{n}g_{m}-2\dot{g}_{n}\partial_{n}g_{m}\right]=2\sqrt{-g}g^{0m}\left[-\dot{g}_{n}\left(\partial_{m}g_{n}-\partial_{n}g_{m}\right)-3\dot{g}_{n}\partial_{n}g_{m}\right].
\label{26}
\end{equation}

{\bf THE FINAL RESULT}:

Hence, according to the expressions (\ref{15}) and (\ref{26}), one
can finally obtain the total contribution to the graviton S-G
coupling as follows
\begin{eqnarray}
\frac{1}{2}K^{2}\sqrt{-g}g^{0m}\left(2\dot{g}_{n}\partial_{m}g_{n}+4\dot{g}_{n}\partial_{n}g_{m}\right)+\frac{1}{4}K^{2}\cdot2\sqrt{-g}g^{0m}\left[-\dot{g}_{n}\left(\partial_{m}g_{n}-\partial_{n}g_{m}\right)-3\dot{g}_{n}\partial_{n}g_{m}\right]                \nonumber \\
=\frac{1}{2}K^{2}\sqrt{-g}g^{0m}\left[-2\dot{g}_{n}\left(\partial_{m}g_{n}-\partial_{n}g_{m}\right)+3\dot{g}_{n}\partial_{m}g_{n}\right].
\end{eqnarray}
Thus one can arrive at
\begin{equation}
{\mathcal
L}_{s-g}=-K^{2}\sqrt{-g}g^{0m}\dot{g}_{n}\left(\partial_{m}g_{n}-\partial_{n}g_{m}\right)=
K^{3}g_{m}\dot{g}_{n}\left(\partial_{m}g_{n}-\partial_{n}g_{m}\right),
\end{equation}
where $\sqrt{-g}g^{0m}=Kh^{0m}=Kg^{m}=-Kg_{m}$ is applied, and the
summation is carried out over the values $ 1, 2, 3$ of the
repeated indices. As stated previously, the Lagrangian density
${\mathcal L}_{s-g}$ is just equal to the Hamiltonian density
describing the interaction of graviton spin with the
gravitomagnetic fields. Hence this note that concerns ourselves
with the detailed mathematical treatment of the coupling of
graviton spin to gravitomagnetic fields may therefore provide us
with a deep insight into the generalized (purely gravitational)
version of Mashhoon's spin-rotation
couplings\cite{Mashhoon2,Mashhoon}. It may be reasonably believed
that such a graviton S-G coupling deserves further detailed
investigation.

\end{document}